\documentclass[aip,jcp]{revtex4-2}

\usepackage{amsmath,amssymb,bm}
\usepackage{mathtools}
\usepackage{graphicx}
\usepackage{booktabs}
\usepackage{hyperref}
\hypersetup{hidelinks}
\usepackage{microtype}

\newcommand{\dd}{\mathrm d}

\begin{document}

\title{Stress Tensor in Fundamental Measure Theory of Hard-Sphere Fluids: A Thermomechanical Approach}

\author{Yury A. Budkov}
\email{ybudkov@hse.ru}
\affiliation{Laboratory of Multiscale Modeling of Molecular Systems, G.~A. Krestov Institute of Solution Chemistry of the Russian Academy of Sciences, Akademicheskaya St. 1, 153045 Ivanovo, Russia}
\affiliation{Laboratory of Computational Physics, HSE University, Tallinskaya St. 34, 123458 Moscow, Russia}
\affiliation{School of Applied Mathematics, HSE University, Tallinskaya St. 34, 123458 Moscow, Russia}

\author{Nikolai N. Kalikin}
\affiliation{Laboratory of Multiscale Modeling of Molecular Systems, G.~A. Krestov Institute of Solution Chemistry of the Russian Academy of Sciences, Akademicheskaya St. 1, 153045 Ivanovo, Russia}

\date{\today}

\begin{abstract}
We derive the local pressure tensor generated by the original Rosenfeld fundamental measure theory (FMT) for hard-sphere fluids from an arbitrary infinitesimal deformation. The variation of a finite-distance FMT kernel is written via the line integrals of the local deformation gradient along the straight segment connecting its endpoints. The resulting symmetric pressure tensor consists of a local term expressed through the standard FMT weighted densities and a finite-range chord term. For a monodisperse hard-sphere fluid, differential relations between the Rosenfeld measures reduce the tensor to a closed one-body form. We demonstrate that no independent pair density or inhomogeneous Ornstein--Zernike equation is required. In the limit of homogeneous fluid the tensor reproduces the Percus--Yevick compressibility equation of state. In planar confinement the normal stress satisfies the hard-wall contact relation, the mechanical surface tension agrees with the surface excess grand potential, and the disjoining pressure agrees with the derivative of the confined grand potential with respect to slit width. The finite-range chord term extends the deformation method from local field theories to nonlocal weighted-density functionals.
\end{abstract}

\maketitle

\section{Introduction}

The local mechanical state of an inhomogeneous fluid is described by the pressure tensor $P_{ij}(\mathbf r)$. Its divergence determines the local force balance, whereas its normal and tangential components determine forces on confining boundaries, disjoining pressures, and interfacial tensions. A density-functional description of both structure and mechanics should generate $P_{ij}$ from the same free-energy functional that determines the equilibrium density.

At the microscopic level, the classical Irving--Kirkwood construction expresses the configurational stress by distributing the force associated with an interacting particle pair along a contour joining the two particles \cite{IrvingKirkwood1950}. Different admissible contours generate different local stress distributions. Schofield and Henderson analyzed this nonuniqueness \cite{SchofieldHenderson1982}, and Shi, et al reviewed its modern interpretation~\cite{ShiSmithSantisoGubbins2023}. The local tangential stress may depend on the contour, but the bulk pressure does not. In planar systems the normal pressure and the integrated normal--tangential stress difference that determines the surface tension are also contour independent. Differences between admissible local stresses form a gauge freedom that preserves force balance and these mechanical observables.

A thermodynamic route follows from the conjugacy between stress and deformation. The pressure tensor is obtained from the response of the free energy to an infinitesimal deformation of space. Mistura formulated this procedure for inhomogeneous fluids in terms of metric variations and derived both the local pressure tensor and the mechanical expression for the surface tension from a free-energy functional~\cite{Mistura1985}. In density-functional and field-theoretical descriptions, equilibrium structure and mechanical response are then generated by the same functional.

For hard-core fluids, Rosenfeld's fundamental measure theory (FMT) writes the excess free energy as a local function of nonlocal weighted densities constructed from geometrical measures of the hard particle~\cite{Rosenfeld1989}. FMT describes packing, wall-induced layering, strong confinement, adsorption, and interfacial thermodynamics. The original Rosenfeld functional reproduces the scaled-particle, or equivalently Percus--Yevick compressibility, equation of state in the homogeneous limit. Later tensorial and White-Bear formulations improve dimensional crossover and bulk thermodynamics~\cite{HansenGoosRoth2006,Roth2010}.

Classical density functional theory (cDFT) has long been used to describe adsorption, capillary phase transitions, and solvent-mediated forces in confined fluids~\cite{Evans1979}. Evans, et al. applied cDFT to capillary condensation and adsorption in slit and cylindrical pores~\cite{EvansMarconiTarazona1986}. For a fluid between parallel walls, Evans and Marini Bettolo Marconi treated the wall separation $H$ as a thermodynamic variable whose conjugate is the solvation force~\cite{EvansMarconi1987}. The same density-functional description generates the equilibrium profile and the derivative of the confined grand potential with respect to geometry.

The corresponding solvation pressure also drives adsorption-induced deformation of nanoporous solids. Density-functional calculations extend continuum adsorption-stress models to nanometer confinement~\cite{GorNeimark2011}. Anisotropic cylindrical pores require distinct pore-level stress components~\cite{BalzerEtAl2017}, while nonconvex pore geometries produce geometry-dependent adsorption stresses and strains~\cite{KolesnikovBudkovGor2020}. Models of adsorption-induced deformation in ordered and disordered porous materials are reviewed in Ref.~\cite{KolesnikovBudkovGor2022Review}. Solvation or disjoining pressure provides the normal fluid-side loading on a pore wall in many systems, while the adsorption-induced surface stress of the solid can contribute separately~\cite{GorKolesnikov2024}. For a nonplanar or anisotropic pore, the fluid loading is specified locally by the traction $t_i=P_{ij}n_j$ rather than by a single scalar pressure. A pressure tensor generated by the same FMT functional as the equilibrium density therefore provides the fluid-side input for adsorption-induced deformation and solid--fluid coupling.

Despite the geometrical structure of FMT, its local mechanical content is not usually written in closed form. Standard FMT calculations provide equilibrium density profiles, grand potentials, adsorption, solvation forces, and surface free energies directly from the one-body functional. Local pressure profiles can be constructed by combining density-functional information with microscopic mechanical relations or pair-correlation expressions~\cite{SunKangZhang2014}, but in this case the stress tensor is not obtained as the direct deformation response of the FMT free energy itself.

Force density functional theory expresses equilibrium through the local
interparticle force density and the Yvon--Born--Green relation%
~\cite{TschoppEtAl2022,SammullerHermannSchmidt2023}. For hard spheres,
FMT can generate the two-body direct correlation function, which is then
combined with the inhomogeneous Ornstein--Zernike equation to obtain pair
correlations and internal forces. This route operates at the two-body
structural level. Our aim is to derive a closed local pressure tensor
directly from the one-body Rosenfeld functional under an arbitrary
virtual deformation of the physical configuration within Euclidean space.

For this purpose, we use the thermomechanical approach%
~\cite{BudkovKalikinBrandyshev2024}, in which stress is obtained from
virtual deformations and spatial symmetries of the thermodynamic
functional. For modified Poisson--Boltzmann theories, the Noether
construction yields stress tensors consistent with the corresponding
Euler--Lagrange equations and expressions for macroscopic and disjoining
forces~\cite{BudkovKolesnikov2022}. For inhomogeneous flexible
polyelectrolyte solutions, deformation of physical Euclidean space gives
the hydrostatic, electrostatic, and polymer conformational contributions
to the stress tensor and the disjoining pressure in a slit nanopore%
~\cite{BudkovKalikin2023}. A covariant Noether formulation was subsequently
developed for field-theoretical free-energy functionals containing scalar
order parameters and their spatial derivatives%
~\cite{BrandyshevBudkov2023} and extended to fluctuating Coulomb fluids%
~\cite{BudkovBrandyshev2023,BrandyshevBudkovJSTAT2023}. Infinitesimal
transformations of physical space were later formulated for scalar and
vector order parameters~\cite{BudkovKalikinBrandyshev2024}; the resulting
normal--tangential stress difference gives the interfacial tension of
electrolyte solutions~\cite{BudkovKalikinBrandyshevSurface2024}. These
results apply to local and gradient free-energy functionals.

The extension of the thermomechanical construction to intrinsically
nonlocal functionals requires an additional step. It has recently been
implemented for inhomogeneous flexible polymer-chain solutions beyond the
ground-state-dominance approximation~\cite{Budkov2026PolymerStress}, where
polymer conformational stress is generated by the nonlocal chain
propagator. Hard-sphere FMT provides a different class of nonlocal
functional, in which the nonlocality is carried by finite-range
geometrical kernels.

For a hard-sphere fluid, the excess free energy in the Rosenfeld and
White-Bear formulations is expressed through weighted densities of the
form
\begin{equation}\label{eq:intro_weighted_density}
 n_{\alpha}(\mathbf r)
 =
 \int \dd^3r'\,
 \rho(\mathbf r')w_{\alpha}(\mathbf r-\mathbf r').
\end{equation}
Each weighted density depends explicitly on the finite separation between
two points of physical space. Under a virtual deformation
\begin{equation}\label{eq:intro_deformation}
 \mathbf X(\mathbf r)=\mathbf r+\mathbf u(\mathbf r),
\end{equation}
this separation changes according to
\begin{equation}\label{eq:intro_distance}
 |\mathbf r-\mathbf r'|
 \longrightarrow
 |\mathbf X(\mathbf r)-\mathbf X(\mathbf r')|.
\end{equation}
The deformation therefore acts on the nonlocal geometrical kernel itself,
whereas for local and gradient functionals it acts on local fields and
their derivatives.

The displacement difference satisfies the exact identity
\begin{equation}\label{eq:intro_chord_identity}
 u_i(\mathbf r)-u_i(\mathbf r')
 =
 (r-r')_j
 \int_0^1 \dd\lambda\,
 \partial_j u_i
 \left[\mathbf r'+\lambda(\mathbf r-\mathbf r')\right].
\end{equation}
Equation~\eqref{eq:intro_chord_identity} converts the variation of a
finite-distance kernel into a local virtual-work form and distributes its
mechanical response along the straight segment joining the two endpoints.
The resulting pressure tensor separates into a local contribution and a
finite-range chord contribution.

The straight chord resembles the contour entering microscopic stress
constructions, but the two constructions have different origins. In
Irving--Kirkwood theory the contour specifies how an intermolecular-force
contribution is distributed in space. Here the straight segment follows
from the variation of the Euclidean distance entering the isotropic FMT
kernel and requires no additional microscopic prescription. Alternative
contour representations differ by divergence-free contributions and may
change local stress profiles while preserving the bulk pressure, the
planar normal pressure, and the integrated surface tension%
~\cite{ShiSmithSantisoGubbins2023}.

For a monodisperse hard-sphere fluid, the Rosenfeld measures satisfy
\begin{equation}\label{eq:intro_measure_relations}
 n_2=\partial_R n_3,
 \qquad
 \mathbf n_2=-\boldsymbol{\nabla}n_3,
\end{equation}
whereas $n_0$, $n_1$, and $\mathbf n_1$ differ from these fields only by
fixed geometrical factors. These relations reduce the deformation problem
to a minimal set of weighted densities. We derive below a closed symmetric
three-dimensional pressure tensor expressed entirely through the one-body
density and the standard FMT fields. No independent pair density or
solution of an inhomogeneous Ornstein--Zernike equation is required.

\section{Stress from a nonlocal weighted-density functional}
\label{sec:general}

\subsection{Virtual deformation of physical space}

We consider an equilibrium fluid described by the intrinsic free-energy functional
\begin{equation}\label{eq:free_energy_split}
 F[\rho]=F_{\rm id}[\rho]+F_{\rm ex}[\rho].
\end{equation}
Unless a boundary or a finite integration interval is written explicitly, three-dimensional spatial integrals in Secs.~\ref{sec:general} and \ref{sec:rosenfeld} are over all of $\mathbb R^3$. We use $\beta=(k_BT)^{-1}$ and the Einstein convention for repeated Cartesian indices.
Throughout the paper we use the pressure-tensor convention: positive diagonal components correspond to compression. The Cauchy stress tensor is therefore $\sigma_{ij}=-P_{ij}$.

The pressure tensor is defined from the response of the intrinsic free energy to an infinitesimal deformation of physical Euclidean space. Here this phrase denotes a virtual deformation of the physical configuration within Euclidean space, not a deformation of the ambient geometry itself. Thus,
\begin{equation}\label{eq:deformation}
 X_i(\mathbf r)=r_i+u_i(\mathbf r),
\end{equation}
where $\mathbf r$ labels the undeformed material point and $\mathbf X$ is its position after deformation. The deformation gradient is
\begin{equation}\label{eq:def_gradient}
 F_{ij}(\mathbf r)
 =
 \frac{\partial X_i}{\partial r_j}
 =
 \delta_{ij}+\partial_j u_i,
\end{equation}
and its Jacobian is $J=\det F$. To first order in the displacement,
\begin{equation}\label{eq:jacobian}
 J=1+\partial_i u_i+O(u^2).
\end{equation}
We define
\begin{equation}\label{eq:Edef}
 E_{ij}\equiv\partial_j u_i
\end{equation}
and the infinitesimal strain tensor
\begin{equation}\label{eq:strain}
 \varepsilon_{ij}=\frac12(E_{ij}+E_{ji}).
\end{equation}

The ambient geometry remains Euclidean: no continuation of the thermodynamic theory to a curved space is assumed. The tensor
\begin{equation}\label{eq:induced_metric}
 g_{ij}=F_{ki}F_{kj}
\end{equation}
will appear below only as a compact representation of the local strain induced by the mapping \eqref{eq:deformation}. It is not treated as an independent metric field.

The number of particles in a material volume element is conserved,
\begin{equation}\label{eq:particle_conservation}
 \rho_u(\mathbf X)\,\dd^3X
 =
 \rho(\mathbf r)\,\dd^3r.
\end{equation}
Hence
\begin{equation}\label{eq:density_transform}
 \rho_u[\mathbf X(\mathbf r)]
 =
 \frac{\rho(\mathbf r)}{J(\mathbf r)}.
\end{equation}
The density entering the deformed functional is therefore fixed by particle conservation and is not varied independently.

The pressure tensor is defined by the virtual-work relation
\begin{equation}\label{eq:virtual_work}
 \delta F
 =
 -\int \dd^3r\,
 P_{ij}(\mathbf r)\partial_j u_i(\mathbf r)
 .
\end{equation}
For a symmetric pressure tensor the same relation can be written as
\begin{equation}\label{eq:virtual_work_strain}
 \delta F
 =
 -\int \dd^3r\,P_{ij}\varepsilon_{ij}.
\end{equation}
For the Rosenfeld functional, this symmetry follows from its rotational structure. The local vector contribution has $b_i\propto q_i$ and is therefore proportional to $q_iq_j$, while the finite-range contribution generated by an isotropic distance kernel is proportional to $e_i e_j$. Hence both anisotropic contributions obtained below are symmetric; symmetry is not imposed as an additional assumption.

\subsection{Ideal-gas contribution}

The ideal contribution in the deformed configuration is written in terms of the thermal wavelength $\Lambda$ as
\begin{equation}\label{eq:ideal_deformed}
 \beta F_{\rm id}^{(u)}
 =
 \int \dd^3X\,
 \rho_u(\mathbf X)
 \left[\ln(\rho_u(\mathbf X)\Lambda^3)-1\right].
\end{equation}
Using Eqs.~\eqref{eq:particle_conservation} and \eqref{eq:density_transform} and returning to material coordinates gives
\begin{equation}\label{eq:ideal_pullback}
 \beta F_{\rm id}^{(u)}
 =
 \int \dd^3r\,
 \rho(\mathbf r)
 \left[
 \ln\frac{\rho(\mathbf r)\Lambda^3}{J(\mathbf r)}-1
 \right].
\end{equation}
To first order, $\delta\ln J=\partial_i u_i$, which gives
\begin{equation}\label{eq:ideal_variation}
 \delta\beta F_{\rm id}
 =
 -\int \dd^3r\,\rho(\mathbf r)\partial_i u_i.
\end{equation}
Comparison with Eq.~\eqref{eq:virtual_work} gives
\begin{equation}\label{eq:ideal_pressure}
 \beta P_{ij}^{\rm id}(\mathbf r)
 =
 \rho(\mathbf r)\delta_{ij}
 .
\end{equation}

\subsection{Scalar nonlocal weighted densities}

Before considering FMT, we establish the structure generated by an isotropic scalar weighted-density functional,
\begin{equation}\label{eq:general_wdf}
 \beta F_{\rm ex}
 =
 \int \dd^3r\,
 \Phi(\{n_\alpha(\mathbf r)\}),
\end{equation}
where
\begin{equation}\label{eq:general_weighted}
 n_\alpha(\mathbf r)
 =
 \int \dd^3r'\,
 \rho(\mathbf r')w_\alpha(|\mathbf r-\mathbf r'|).
\end{equation}
The kernels $w_\alpha(s)$ may be ordinary functions or distributions. In FMT they include Heaviside and Dirac delta functions.

Under the deformation \eqref{eq:deformation}, the Euclidean distance between two material points becomes
\begin{equation}\label{eq:deformed_distance}
 S(\mathbf r,\mathbf r')
 =
 |\mathbf X(\mathbf r)-\mathbf X(\mathbf r')|.
\end{equation}
Using particle conservation, the deformed weighted density pulled back to the reference coordinates reads
\begin{equation}\label{eq:weighted_pullback}
 n_\alpha^{(u)}(\mathbf r)
 =
 \int \dd^3r'\,
 \rho(\mathbf r')
 w_\alpha[S(\mathbf r,\mathbf r')].
\end{equation}
Here the square brackets in $w_\alpha[S(\mathbf r,\mathbf r')]$ denote ordinary evaluation of the scalar function $w_\alpha$ at the deformed distance $S$; they do not denote a functional argument. There is no Jacobian under the convolution integral because the factor from the deformed volume element is exactly cancelled by the density transformation.

Define
\begin{equation}\label{eq:s_def}
 \mathbf s=\mathbf r-\mathbf r',
 \qquad
 s=|\mathbf s|,
 \qquad
 \mathbf e=\frac{\mathbf s}{s}.
\end{equation}
The first-order variation of the distance is
\begin{equation}\label{eq:deltaS1}
 \delta S
 =
 e_i[u_i(\mathbf r)-u_i(\mathbf r')].
\end{equation}
The displacement difference is represented exactly as
\begin{equation}\label{eq:chord_identity}
 u_i(\mathbf r)-u_i(\mathbf r')
 =
 s_j\int_0^1\dd\lambda\,
 \partial_j u_i(\mathbf r'+\lambda\mathbf s).
\end{equation}
Consequently,
\begin{equation}\label{eq:deltaS_chord}
 \delta S
 =
 s e_i e_j
 \int_0^1\dd\lambda\,
 E_{ij}(\mathbf r'+\lambda\mathbf s)
 .
\end{equation}
Equation~\eqref{eq:deltaS_chord} represents the finite-distance variation as an integral of the local deformation gradient along the straight segment joining the two endpoints.

The variation of Eq.~\eqref{eq:weighted_pullback} is
\begin{equation}\label{eq:delta_n_general1}
 \delta n_\alpha(\mathbf r)
 =
 \int \dd^3r'\,\rho(\mathbf r')w_\alpha'(s)\delta S,
\end{equation}
where $w_\alpha'(s)\equiv \dd w_\alpha(s)/\dd s$ denotes differentiation with respect to the scalar argument $s$. Equivalently,
\begin{equation}\label{eq:delta_n_general2}
\begin{aligned}
 \delta n_\alpha(\mathbf r)
 ={}&
 \int \dd^3r'\,\rho(\mathbf r')s w_\alpha'(s)e_i e_j
 \int_0^1\dd\lambda\,
 E_{ij}(\mathbf r'+\lambda\mathbf s).
\end{aligned}
\end{equation}
Introduce
\begin{equation}\label{eq:psi_general}
 \psi_\alpha(\mathbf r)
 =
 \frac{\partial\Phi}{\partial n_\alpha(\mathbf r)}.
\end{equation}
After pullback, the excess free energy is
\begin{equation}\label{eq:general_F_pullback}
 \beta F_{\rm ex}^{(u)}
 =
 \int \dd^3r\,J(\mathbf r)
 \Phi(\{n_\alpha^{(u)}\}).
\end{equation}
Its first variation is
\begin{equation}\label{eq:general_F_variation}
 \delta\beta F_{\rm ex}
 =
 \int \dd^3r\,\Phi E_{kk}
 +
 \sum_\alpha\int \dd^3r\,\psi_\alpha\delta n_\alpha.
\end{equation}
Substitution of Eq.~\eqref{eq:delta_n_general2} gives the local virtual-work form
\begin{equation}\label{eq:general_virtual_work}
 \delta\beta F_{\rm ex}
 =
 \int \dd^3z\,
 [\Phi(\mathbf z)\delta_{ij}-C_{ij}^{(s)}(\mathbf z)]E_{ij}(\mathbf z),
\end{equation}
with
\begin{equation}\label{eq:general_chord_stress}
\begin{aligned}
 C_{ij}^{(s)}(\mathbf z)
 ={}&
 -\sum_\alpha
 \int \dd^3r\,\dd^3r'\,
 \rho(\mathbf r')\psi_\alpha(\mathbf r)
 s w_\alpha'(s)e_i e_j\\
 &\times
 \int_0^1\dd\lambda\,
 \delta^{(3)}[\mathbf z-\mathbf r'-\lambda\mathbf s].
\end{aligned}
\end{equation}
The relation of the straight-chord representation to the usual contour freedom of microscopic stress is given in Appendix~\ref{app:general_chord}.

Combining Eqs.~\eqref{eq:ideal_pressure} and \eqref{eq:general_virtual_work}, the pressure tensor of the scalar weighted-density theory is
\begin{equation}\label{eq:general_pressure}
 \beta P_{ij}
 =
 (\rho-\Phi)\delta_{ij}+C_{ij}^{(s)}
 .
\end{equation}

\subsection{Mechanical equilibrium}

Consider the grand potential
\begin{equation}\label{eq:grand_potential}
 \Omega[\rho]
 =
 F[\rho]
 +\int \dd^3r\,\rho V_{\rm ext}
 -\mu\int \dd^3r\,\rho.
\end{equation}
The particle-number term is invariant under a material deformation. The external-field contribution becomes
\begin{equation}\label{eq:Fext_pullback}
 F_{\rm ext}^{(u)}
 =
 \int \dd^3r\,\rho(\mathbf r)
 V_{\rm ext}[\mathbf X(\mathbf r)],
\end{equation}
so that
\begin{equation}\label{eq:Fext_variation}
 \delta F_{\rm ext}
 =
 \int \dd^3r\,\rho(\mathbf r)u_i(\mathbf r)
 \partial_iV_{\rm ext}(\mathbf r).
\end{equation}
At equilibrium, $\delta\Omega=0$. Using Eq.~\eqref{eq:virtual_work} and integrating by parts for a displacement that vanishes at the outer boundary gives
\begin{equation}\label{eq:force_balance_pre}
 \int \dd^3r\,u_i
 [\partial_jP_{ij}+\rho\partial_iV_{\rm ext}]=0.
\end{equation}
Since $\mathbf u(\mathbf r)$ is arbitrary,
\begin{equation}\label{eq:force_balance}
 \partial_jP_{ij}(\mathbf r)
 =
 -\rho(\mathbf r)\partial_iV_{\rm ext}(\mathbf r)
 .
\end{equation}

\section{Rosenfeld fundamental measure theory}
\label{sec:rosenfeld}

\subsection{Functional and independent fundamental measures}

For a monodisperse hard-sphere fluid of radius $R$, the original Rosenfeld excess free-energy functional is
\begin{equation}\label{eq:R_Fex}
 \beta F_{\rm ex}=\int \dd^3r\,\Phi(\mathbf r),
\end{equation}
with
\begin{equation}\label{eq:R_Phi_original}
\begin{aligned}
 \Phi={}&
 -n_0\ln(1-n_3)
 +\frac{n_1n_2-\mathbf n_1\cdot\mathbf n_2}{1-n_3}+\frac{n_2^3-3n_2\mathbf n_2^2}{24\pi(1-n_3)^2}.
\end{aligned}
\end{equation}
The scalar weights are
\begin{equation}\label{eq:R_scalar_weights1}
 w_3(\mathbf s)=\Theta(R-s),
 \qquad
 w_2(\mathbf s)=\delta(R-s),
\end{equation}
\begin{equation}\label{eq:R_scalar_weights2}
 w_1=\frac{w_2}{4\pi R},
 \qquad
 w_0=\frac{w_2}{4\pi R^2},
\end{equation}
and the vector weights are
\begin{equation}\label{eq:R_vector_weights}
 \mathbf w_2(\mathbf s)=\mathbf e\,\delta(R-s),
 \qquad
 \mathbf w_1=\frac{\mathbf w_2}{4\pi R}.
\end{equation}
The weighted densities are
\begin{equation}\label{eq:R_scalar_densities}
 n_\alpha(\mathbf r)
 =
 \int \dd^3r'\,\rho(\mathbf r')w_\alpha(\mathbf r-\mathbf r'),
\end{equation}
\begin{equation}\label{eq:R_vector_densities}
 \mathbf n_a(\mathbf r)
 =
 \int \dd^3r'\,\rho(\mathbf r')\mathbf w_a(\mathbf r-\mathbf r'),
 \qquad a=1,2.
\end{equation}

For a monodisperse fluid only three independent geometrical fields are required. From the definitions,
\begin{equation}\label{eq:n2_from_n3}
 n_2=\partial_R n_3,
\end{equation}
with
\begin{equation}\label{eq:n3_def}
 n_3(\mathbf r)
 =
 \int \dd^3r'\,\rho(\mathbf r')
 \Theta(R-|\mathbf r-\mathbf r'|),
\end{equation}
and
\begin{equation}\label{eq:q_from_n3}
 \mathbf n_2=-\boldsymbol\nabla n_3.
\end{equation}
Furthermore,
\begin{equation}\label{eq:dependent_measures}
 n_1=\frac{n_2}{4\pi R},
 \qquad
 n_0=\frac{n_2}{4\pi R^2},
 \qquad
 \mathbf n_1=\frac{\mathbf n_2}{4\pi R}.
\end{equation}
The proof of Eqs.~\eqref{eq:n2_from_n3}--\eqref{eq:dependent_measures}, including their behavior under deformation, is given in Appendix~\ref{app:rosenfeld_vector}.

Introduce
\begin{equation}\label{eq:q_D_def}
 \mathbf q\equiv\mathbf n_2,
 \qquad
 q^2=\mathbf q^2,
 \qquad
 D\equiv1-n_3.
\end{equation}
Equation~\eqref{eq:R_Phi_original} can then be written exactly as
\begin{equation}\label{eq:R_Phi_reduced}
\begin{aligned}
 \Phi(n_3,n_2,q^2)={}&
 -\frac{n_2}{4\pi R^2}\ln D
 +\frac{n_2^2-q^2}{4\pi RD}+\frac{n_2^3-3n_2q^2}{24\pi D^2}.
\end{aligned}
\end{equation}
Define
\begin{equation}\label{eq:a_chi_b_def}
 a\equiv\frac{\partial\Phi}{\partial n_3},
 \qquad
 \chi\equiv\frac{\partial\Phi}{\partial n_2},
 \qquad
 b_i\equiv\frac{\partial\Phi}{\partial q_i}.
\end{equation}
Direct differentiation gives
\begin{equation}\label{eq:a_explicit}
 a
 =
 \frac{n_2}{4\pi R^2D}
 +\frac{n_2^2-q^2}{4\pi RD^2}
 +\frac{n_2^3-3n_2q^2}{12\pi D^3},
\end{equation}
\begin{equation}\label{eq:chi_explicit}
 \chi
 =
 -\frac{\ln D}{4\pi R^2}
 +\frac{n_2}{2\pi RD}
 +\frac{n_2^2-q^2}{8\pi D^2},
\end{equation}
and
\begin{equation}\label{eq:b_explicit}
 b_i
 =
 -\left[
 \frac{1}{2\pi RD}
 +\frac{n_2}{4\pi D^2}
 \right]q_i.
\end{equation}

\subsection{Deformation of the vector fundamental measure}

The scalar measures $n_3$ and $n_2$ vary through the deformed distance. The vector measure also changes orientation. For the Rosenfeld functional its variation follows from the exact relation \eqref{eq:q_from_n3}.

In the deformed physical coordinates,
\begin{equation}\label{eq:q_deformed_phys}
 \mathbf q^{(u)}=-\boldsymbol\nabla_X n_3^{(u)}.
\end{equation}
Define the material covector
\begin{equation}\label{eq:Q_def}
 Q_i(\mathbf r)=-\partial_i n_3^{(u)}(\mathbf r).
\end{equation}
By the chain rule,
\begin{equation}\label{eq:q_transform}
 Q_i=F_{ai}q_a^{(u)},
 \qquad
 q_a^{(u)}=(F^{-T})_{ai}Q_i,
\end{equation}
where $F^{-T}\equiv(F^{-1})^T=(F^T)^{-1}$. Hence the physical magnitude is
\begin{equation}\label{eq:q2_metric}
 (q^{(u)})^2=g^{ij}Q_iQ_j,
\end{equation}
where $g^{ij}$ is the inverse of the induced Euclidean tensor $g_{ij}$ defined in Eq.~\eqref{eq:induced_metric}.

At zero deformation, $Q_i=q_i$. To first order,
\begin{equation}\label{eq:delta_Q_g}
 \delta Q_i=-\partial_i\delta n_3,
 \qquad
 \delta g^{ij}=-(E_{ij}+E_{ji}),
\end{equation}
and therefore
\begin{equation}\label{eq:delta_q2}
 \delta q^2
 =
 -2q_iq_jE_{ij}
 -2q_i\partial_i\delta n_3.
\end{equation}
Since
\begin{equation}\label{eq:b_chain}
 b_i=2\frac{\partial\Phi}{\partial q^2}q_i,
\end{equation}
the vector contribution becomes
\begin{equation}\label{eq:vector_variation1}
 \frac{\partial\Phi}{\partial q^2}\delta q^2
 =
 -b_iq_jE_{ij}-b_i\partial_i\delta n_3.
\end{equation}
After integration by parts,
\begin{equation}\label{eq:vector_integration_parts}
 -\int \dd^3r\,b_i\partial_i\delta n_3
 =
 \int \dd^3r\,(\partial_i b_i)\delta n_3,
\end{equation}
where boundary terms are understood in the weak/distributional sense and vanish at the outer boundary of the full support of the weighted fields. Define
\begin{equation}\label{eq:A_def}
 A(\mathbf r)=a(\mathbf r)+\partial_i b_i(\mathbf r).
\end{equation}
Appendix~\ref{app:rosenfeld_vector} verifies Eq.~\eqref{eq:A_def} by varying the original vector weight.

\subsection{Full three-dimensional pressure tensor}

Under deformation,
\begin{equation}\label{eq:n3_deformed}
 n_3^{(u)}(\mathbf r)
 =
 \int \dd^3r'\,\rho(\mathbf r')
 \Theta[R-S(\mathbf r,\mathbf r')].
\end{equation}
Equations~\eqref{eq:deltaS_chord} and \eqref{eq:n3_deformed} give
\begin{equation}\label{eq:delta_n3}
\begin{aligned}
 \delta n_3(\mathbf r)
 ={}&
 -\int \dd^3r'\,\rho(\mathbf r')
 s\delta(R-s)e_i e_j
 \int_0^1\dd\lambda\,
 E_{ij}(\mathbf r'+\lambda\mathbf s),
\end{aligned}
\end{equation}
and, using $n_2=\partial_Rn_3$,
\begin{equation}\label{eq:delta_n2}
\begin{aligned}
 \delta n_2(\mathbf r)
 ={}&
 -\int \dd^3r'\,\rho(\mathbf r')
 s\delta'(R-s)e_i e_j
 \int_0^1\dd\lambda\,
 E_{ij}(\mathbf r'+\lambda\mathbf s).
\end{aligned}
\end{equation}
Here and below
\begin{equation}\label{eq:delta_prime_def}
 \delta'(R-s)
 \equiv
 \left.\frac{\dd\delta(\xi)}{\dd\xi}\right|_{\xi=R-s}
 =\partial_R\delta(R-s).
\end{equation}

After pullback,
\begin{equation}\label{eq:R_F_pullback}
 \beta F_{\rm ex}^{(u)}
 =
 \int \dd^3r\,J(\mathbf r)
 \Phi[n_3^{(u)},n_2^{(u)},(q^{(u)})^2].
\end{equation}
Using Eqs.~\eqref{eq:vector_variation1}--\eqref{eq:A_def}, its first variation can be written as
\begin{equation}\label{eq:R_F_variation_compact}
\begin{aligned}
 \delta\beta F_{\rm ex}
 ={}&
 \int \dd^3r\,[\Phi\delta_{ij}-b_iq_j]E_{ij}\\
 &+\int \dd^3r\,[A\delta n_3+\chi\delta n_2].
\end{aligned}
\end{equation}
Substituting Eqs.~\eqref{eq:delta_n3} and \eqref{eq:delta_n2} yields
\begin{equation}\label{eq:R_F_variation_final}
 \delta\beta F_{\rm ex}
 =
 \int \dd^3z\,[\Phi\delta_{ij}-b_iq_j-C_{ij}]E_{ij}(\mathbf z),
\end{equation}
where
\begin{equation}\label{eq:C_two_center}
\begin{aligned}
 C_{ij}(\mathbf z)
 ={}&
 \int \dd^3r\,\dd^3r'\,
 \rho(\mathbf r')s e_i e_j\\
 &\times
 [A(\mathbf r)\delta(R-s)+\chi(\mathbf r)\delta'(R-s)]\\
 &\times
 \int_0^1\dd\lambda\,
 \delta^{(3)}[\mathbf z-\mathbf r'-\lambda(\mathbf r-\mathbf r')].
\end{aligned}
\end{equation}
We refer to $C_{ij}$ as the chord contribution. A particle at $\mathbf r'$ contributes to a fundamental measure centered at $\mathbf r$, and the deformation response of this finite separation is distributed along the straight segment joining the two points. The $\delta(R-s)$ part originates from the volume measure, whereas the $\delta'(R-s)$ part reflects the relation $n_2=\partial_Rn_3$.

Combining Eqs.~\eqref{eq:ideal_pressure} and \eqref{eq:R_F_variation_final} with the virtual-work definition gives
\begin{equation}\label{eq:full_pressure_compact}
 \beta P_{ij}(\mathbf r)
 =
 [\rho(\mathbf r)-\Phi(\mathbf r)]\delta_{ij}
 +b_i(\mathbf r)q_j(\mathbf r)
 +C_{ij}(\mathbf r)
 .
\end{equation}
Using Eq.~\eqref{eq:b_explicit},
\begin{equation}\label{eq:full_pressure_explicit}
\begin{aligned}
 \beta P_{ij}
 ={}&
 (\rho-\Phi)\delta_{ij}+C_{ij}\\
 &-
 \left[
 \frac{1}{2\pi R(1-n_3)}
 +\frac{n_2}{4\pi(1-n_3)^2}
 \right]q_iq_j .
\end{aligned}
\end{equation}
Both anisotropic terms are symmetric because $q_iq_j=q_jq_i$ and the chord kernel is proportional to $e_i e_j$. Hence
\begin{equation}\label{eq:pressure_symmetric}
 P_{ij}=P_{ji}.
\end{equation}

A one-center representation follows from setting $\mathbf s=\mathbf r-\mathbf r'$ in Eq.~\eqref{eq:C_two_center} and performing the line-delta integration:
\begin{equation}\label{eq:C_one_center}
\begin{aligned}
 C_{ij}(\mathbf z)
 ={}&
 \int_0^1\dd\lambda\int \dd^3s\,
 s e_i e_j\,
 \rho(\mathbf z-\lambda\mathbf s)\\
 &\times\Big\{
 A[\mathbf z+(1-\lambda)\mathbf s]\delta(R-s)\\
 &\qquad
 +\chi[\mathbf z+(1-\lambda)\mathbf s]\delta'(R-s)
 \Big\}.
\end{aligned}
\end{equation}

For smooth profiles the radial distribution derivative can be eliminated. For a test function $f(s)$,
\begin{equation}\label{eq:delta_prime_identity}
 \int_0^\infty \dd s\,f(s)\delta'(R-s)
 =
 \left.\frac{\partial f(s)}{\partial s}\right|_{s=R},
\end{equation}
where the derivative acts on the integration variable $s$ with the hard-sphere radius and the FMT fields held fixed; it is not a total derivative with respect to $R$. For a smooth density profile,
\begin{equation}\label{eq:C_smooth}
\begin{aligned}
 C_{ij}(\mathbf r)
 ={}&
 \int_0^1\dd\lambda\int \dd\Omega_{\mathbf e}\,e_i e_j
 \Bigg\{
 R^3\rho_-A_+\\
 &\quad+
 \left.\frac{\partial}{\partial s}
 \left[
 s^3\rho(\mathbf r-\lambda s\mathbf e)
 \chi(\mathbf r+(1-\lambda)s\mathbf e)
 \right]\right|_{s=R}
 \Bigg\},
\end{aligned}
\end{equation}
where
\begin{equation}\label{eq:rhoApm}
 \rho_-=\rho(\mathbf r-\lambda R\mathbf e),
 \qquad
 A_+=A[\mathbf r+(1-\lambda)R\mathbf e].
\end{equation}
Writing also $\chi_+=\chi[\mathbf r+(1-\lambda)R\mathbf e]$, Eq.~\eqref{eq:C_smooth} can be expanded as
\begin{equation}\label{eq:C_smooth_expanded}
\begin{aligned}
 C_{ij}(\mathbf r)
 ={}&\int_0^1\dd\lambda\int\dd\Omega_{\mathbf e}\,e_i e_j
 \Big\{
 R^3\rho_-A_+ +3R^2\rho_-\chi_+\\
 &\quad-\lambda R^3(\mathbf e\cdot\boldsymbol\nabla\rho_-)\chi_+\\
 &\quad+(1-\lambda)R^3\rho_-
 (\mathbf e\cdot\boldsymbol\nabla\chi_+)
 \Big\}.
\end{aligned}
\end{equation}
For a hard-wall density profile, Eq.~\eqref{eq:C_one_center} is the safer starting point because the density has a jump at the center-accessible boundary. The distributional treatment is discussed in Appendix~\ref{app:planar}.

\subsection{Homogeneous limit}

For a uniform fluid, $\rho(\mathbf r)=\rho_b$, the vector measures vanish,
\begin{equation}\label{eq:bulk_vector_zero}
 \mathbf q=\mathbf b=0,
\end{equation}
and $A=a$. Equation~\eqref{eq:C_one_center} then gives
\begin{equation}\label{eq:C_bulk1}
 C_{ij}^{\rm bulk}
 =
 \rho_b\int \dd^3s\,s e_i e_j
 [a\delta(R-s)+\chi\delta'(R-s)].
\end{equation}
Using
\begin{equation}\label{eq:angular_identity}
 \int \dd\Omega\,e_i e_j=\frac{4\pi}{3}\delta_{ij},
\end{equation}
we obtain
\begin{equation}\label{eq:C_bulk2}
 C_{ij}^{\rm bulk}
 =
 (n_3a+n_2\chi)\delta_{ij}.
\end{equation}
Therefore
\begin{equation}\label{eq:bulk_pressure_general}
 \beta p
 =
 \rho_b-\Phi_b+n_3a+n_2\chi
 .
\end{equation}
Here $\Phi_b$ denotes the Rosenfeld excess free-energy density $\Phi$ evaluated at the homogeneous bulk weighted densities.
For a monodisperse fluid,
\begin{equation}\label{eq:bulk_measures}
 n_3=\eta,
 \qquad
 n_2=4\pi R^2\rho_b=\frac{3\eta}{R},
 \qquad
 \eta=\frac{4\pi}{3}R^3\rho_b.
\end{equation}
Substitution of Eqs.~\eqref{eq:R_Phi_reduced}, \eqref{eq:a_explicit}, and \eqref{eq:chi_explicit} into Eq.~\eqref{eq:bulk_pressure_general} gives
\begin{equation}\label{eq:PY_EOS}
 \frac{\beta p}{\rho_b}
 =
 \frac{1+\eta+\eta^2}{(1-\eta)^3}
 .
\end{equation}
Equation~\eqref{eq:PY_EOS} reproduces exactly the Percus--Yevick compressibility equation of state of the original Rosenfeld functional.

\section{Planar geometry}
\label{sec:planar}

\subsection{One-dimensional weighted densities}

We consider a system translationally invariant in the $x$ and $y$ directions,
\begin{equation}\label{eq:planar_density}
 \rho(\mathbf r)=\rho(z).
\end{equation}
All scalar weighted densities depend only on $z$, whereas the vector surface measure has the form
\begin{equation}\label{eq:planar_q}
 \mathbf q(z)=q(z)\mathbf e_z.
\end{equation}
The three-dimensional convolutions reduce to
\begin{equation}\label{eq:1D_scalar_conv}
 n_\alpha(z)=\int_{-\infty}^{\infty} \dd z'\,\rho(z')W_\alpha(z-z'),
\end{equation}
\begin{equation}\label{eq:1D_vector_conv}
 q(z)=\int_{-\infty}^{\infty} \dd z'\,\rho(z')V_2(z-z').
\end{equation}
For a hard slit with center-accessible interval $0<z<H$, $\rho(z)=0$ outside this interval, and the limits in Eqs.~\eqref{eq:1D_scalar_conv} and \eqref{eq:1D_vector_conv} may therefore be replaced by $0$ and $H$. Writing $h=z-z'$, the kernels are
\begin{equation}\label{eq:W3}
 W_3(h)=\pi(R^2-h^2)\Theta(R-|h|),
\end{equation}
\begin{equation}\label{eq:W2}
 W_2(h)=2\pi R\Theta(R-|h|),
\end{equation}
\begin{equation}\label{eq:W10}
 W_1(h)=\frac12\Theta(R-|h|),
 \qquad
 W_0(h)=\frac{1}{2R}\Theta(R-|h|),
\end{equation}
and
\begin{equation}\label{eq:V21}
 V_2(h)=2\pi h\Theta(R-|h|),
 \qquad
 V_1(h)=\frac{h}{2R}\Theta(R-|h|).
\end{equation}
Their derivation and normalization are given in Appendix~\ref{app:planar}.

\subsection{Normal and tangential components}

Planar symmetry implies
\begin{equation}\label{eq:PNPT_def}
 P_{xx}=P_{yy}\equiv P_T,
 \qquad
 P_{zz}\equiv P_N.
\end{equation}
Equation~\eqref{eq:full_pressure_compact} gives
\begin{equation}\label{eq:PN}
 \beta P_N(z)
 =
 \rho(z)-\Phi(z)+q(z)b(z)+C_{zz}(z)
 ,
\end{equation}
where
\begin{equation}\label{eq:b_planar}
 b(z)
 =
 -\left[
 \frac{1}{2\pi R(1-n_3)}
 +\frac{n_2}{4\pi(1-n_3)^2}
 \right]q(z),
\end{equation}
and
\begin{equation}\label{eq:PT}
 \beta P_T(z)
 =
 \rho(z)-\Phi(z)+C_{xx}(z)
 .
\end{equation}

For a smooth profile, introduce $u=\cos\theta$, where $\theta$ is the angle between the chord direction and the $z$ axis, and define
\begin{equation}\label{eq:G_def}
\begin{aligned}
 \mathcal G(z,u,\lambda)={}&
 R^3\rho(z-\lambda Ru)
 A[z+(1-\lambda)Ru]\\
 &+\left.\frac{\partial}{\partial s}
 \Big\{
 s^3\rho(z-\lambda su)
 \chi[z+(1-\lambda)su]
 \Big\}\right|_{s=R}.
\end{aligned}
\end{equation}
Then
\begin{equation}\label{eq:Czz}
 C_{zz}(z)
 =
 2\pi\int_0^1\dd\lambda\int_{-1}^{1}\dd u\,u^2\mathcal G(z,u,\lambda),
\end{equation}
and
\begin{equation}\label{eq:Cxx}
 C_{xx}(z)
 =
 \pi\int_0^1\dd\lambda\int_{-1}^{1}\dd u\,(1-u^2)\mathcal G(z,u,\lambda).
\end{equation}
Subtracting the tangential component from the normal component gives
\begin{equation}\label{eq:stress_anisotropy}
\begin{aligned}
 \beta[P_N(z)-P_T(z)]={}&q(z)b(z)\\
 &+\pi\int_0^1\dd\lambda\int_{-1}^{1}\dd u\,
 (3u^2-1)\\
 &\qquad\times\mathcal G(z,u,\lambda).
\end{aligned}
\end{equation}
For hard walls, the equivalent distributional representation in Appendix~\ref{app:planar} is used instead of differentiating the discontinuous density profile.

\subsection{Mechanical equilibrium and the hard-wall contact theorem}

If $V_{\rm ext}=V_{\rm ext}(z)$, Eq.~\eqref{eq:force_balance} reduces to
\begin{equation}\label{eq:planar_force_balance}
 \frac{\dd P_N}{\dd z}
 =
 -\rho(z)\frac{\dd V_{\rm ext}}{\dd z}
 .
\end{equation}
In any region without an external body force,
\begin{equation}\label{eq:PN_const}
 P_N(z)=\text{const}.
\end{equation}
For a single wall with a homogeneous bulk region, this constant is the bulk pressure $p_b$. In a finite slit, the constant is the slit normal pressure and generally differs from $p_b$ by the disjoining pressure.

Consider a hard wall and let $z=0$ denote the plane accessible to particle centers,
\begin{equation}\label{eq:wall_accessibility}
 \rho(z)=0,\qquad z<0.
\end{equation}
A virtual displacement of the hard boundary gives the standard contact relation
\begin{equation}\label{eq:contact_theorem}
 \rho(0^+)=\beta P_N(0^+)=\beta p_b
 .
\end{equation}
For a nonlocal functional, the support of the excess free-energy density is larger than the region accessible to particle centers. Although $\rho(z)=0$ for $z<0$, the Rosenfeld weighted densities can remain nonzero down to $z=-R$. The integral of $\Phi$ must include this entire layer. Truncating the free-energy integration at the center-accessible plane violates the translational consistency required for Eq.~\eqref{eq:contact_theorem}.

\subsection{Surface tension}

The mechanical surface tension of a planar interface is
\begin{equation}\label{eq:gamma_mech}
 \gamma
 =
 \int_{-\infty}^{\infty} \dd z\,[P_N(z)-P_T(z)]
 .
\end{equation}
For a fluid interface the limits in Eq.~\eqref{eq:gamma_mech} are the two homogeneous bulk regions. For a single hard wall with the center-accessible plane at $z=0$, the same integral may be written as $\int_{-R}^{\infty}\dd z$ because the FMT stress anisotropy vanishes for $z<-R$. The value obtained in this convention is the surface excess $\gamma_c$ referred to the center-accessible plane; the shift to the physical wall is given below.

Consider the volume-preserving affine deformation
\begin{equation}\label{eq:area_deformation}
 X=(1+\epsilon)x,
 \qquad
 Y=(1+\epsilon)y,
 \qquad
 Z=(1-2\epsilon)z.
\end{equation}
To first order, $\delta V=0$ and $\delta A/A=2\epsilon$. The virtual work is
\begin{equation}\label{eq:gamma_virtual_work}
 \delta\Omega
 =
 2\epsilon A\int_{-\infty}^{\infty} \dd z\,[P_N(z)-P_T(z)],
\end{equation}
whereas thermodynamics gives $\delta\Omega=\gamma\delta A=2\epsilon A\gamma$. Equating the two expressions gives Eq.~\eqref{eq:gamma_mech}.

For the chord contribution, integration over $z$ removes the explicit chord-parameter dependence. In Eq.~\eqref{eq:C_one_center}, set
\begin{equation}\label{eq:zp_substitution}
 z_p=z-\lambda su.
\end{equation}
The second endpoint becomes $z_p+su$, so the integrand loses its explicit $\lambda$ dependence and $\int_0^1\dd\lambda=1$. For a smooth profile,
\begin{equation}\label{eq:gamma_compact}
\begin{aligned}
 \beta\gamma={}&
 \int_{-\infty}^{\infty} \dd z\,q(z)b(z)\\
 &+\pi\int_{-1}^{1}\dd u\,(3u^2-1)\\
 &\quad\times
 \int_{-\infty}^{\infty} \dd z\,\rho(z)
 \left[
 R^3A(z+Ru)+3R^2\chi(z+Ru)
 \right.\\
 &\left.
 \hspace{29mm}
 +R^3u\chi'(z+Ru)
 \right],
\end{aligned}
\end{equation}
where $\chi'(z)=\dd\chi(z)/\dd z$. For a single hard wall the first integral has support only for $z\ge -R$, while the density in the second integral restricts it to $z\ge0$. Equation~\eqref{eq:gamma_compact} remains valid for the hard-wall profile if it is derived from the distributional chord form by carrying out the $z$ integration before the radial $s$ integration; no derivative of the discontinuous density is then required. The local hard-wall representation and this integrated form are derived in Appendix~\ref{app:planar}.

For a hard wall, the dividing surface must be specified. If $z=0$ is the contact plane of particle centers, Eq.~\eqref{eq:gamma_mech} defines $\gamma_c$ relative to that plane. The physical wall surface lies one particle radius to the left, $z_{\rm wall}=-R$. Moving the dividing surface from the contact plane to the physical wall increases the reference bulk volume per unit area by $R$; therefore
\begin{equation}\label{eq:gamma_shift}
 \gamma_{\rm hw}=\gamma_c+p_bR
 .
\end{equation}

The thermodynamic surface free energy is obtained from
\begin{equation}\label{eq:gamma_thermo_def}
 \gamma=\frac{\Omega+p_bV}{A}
\end{equation}
using the same dividing surface. For a single hard wall with the center-accessible plane at $z=0$,
\begin{equation}\label{eq:gamma_thermo}
\begin{aligned}
 \beta\gamma_c^{\rm th}={}&
 \int_{-R}^{0}\dd z\,\Phi(z)\\
 &+\int_0^\infty\dd z\,
 \Big\{
 \rho(z)[\ln(\rho(z)\Lambda^3)-1-\beta\mu]\\
 &\hspace{32mm}+\Phi(z)+\beta p_b
 \Big\}.
\end{aligned}
\end{equation}
The equality of the mechanical and thermodynamic routes,
\begin{equation}\label{eq:gamma_equality}
 \gamma_c^{\rm mech}=\gamma_c^{\rm th},
\end{equation}
is a stringent internal consistency test. For a given approximate free-energy functional, the equality must hold when the mechanical stress is derived consistently from that same functional. It therefore tests the stress construction and its numerical implementation, rather than the accuracy of the Rosenfeld approximation relative to the exact hard-sphere fluid.

\section{Numerical tests}
\label{sec:numerical}

We test the stress tensor against independent thermodynamic relations in planar geometries. All calculations use the original monodisperse Rosenfeld functional. We set $R=1$, $\sigma=2R$, $\beta=1$, and $\Lambda=1$; the latter choice only fixes the additive reference of the chemical potential. The slit width $H$ denotes the distance between the two contact planes accessible to particle centers. The physical wall-to-wall distance is therefore $H+2R$.

Inside the accessible interval, the equilibrium profile satisfies
\begin{equation}\label{eq:numerics_EL}
 \ln[\rho(z)\Lambda^3]+c_{\rm ex}(z)=\beta\mu,
 \qquad 0<z<H,
\end{equation}
with $\rho(z)=0$ for $z\le0$ and $z\ge H$. Here
\begin{equation}\label{eq:numerics_cex}
\begin{aligned}
 c_{\rm ex}(z)
 &\equiv\frac{\delta\beta F_{\rm ex}}{\delta\rho(z)}\\
 &=\int_{-\infty}^{\infty}\dd z'\,
 \Big\{
 a(z')W_3(z'-z)+\chi(z')W_2(z'-z)\\
 &\hspace{35mm}+b(z')V_2(z'-z)
 \Big\}.
\end{aligned}
\end{equation}
The odd parity of $V_2$ gives the equivalent convolution form with $-b(z')V_2(z-z')$. The reservoir chemical potential was evaluated from the homogeneous FMT solution,
\begin{equation}\label{eq:numerics_mu_bulk}
 \beta\mu
 =\ln(\rho_b\Lambda^3)
 +\frac{4\pi R^3}{3}a_b+4\pi R^2\chi_b,
\end{equation}
and $p_b$ was obtained from Eq.~\eqref{eq:PY_EOS}.

For a slit, the grand potential per unit area was evaluated as
\begin{equation}\label{eq:numerics_grand_potential}
\begin{aligned}
 \frac{\beta\Omega}{A}
 ={}&\int_0^H\dd z\,\rho(z)
 \left\{\ln[\rho(z)\Lambda^3]-1-\beta\mu\right\}\\
 &+\int_{-R}^{H+R}\dd z\,\Phi(z).
\end{aligned}
\end{equation}
The two contributions have different supports. The ideal-gas integrand may formally be extended outside $0<z<H$ because $\rho$ vanishes there, whereas the weighted fields, and hence $\Phi(z)$, remain nonzero within one radius beyond each contact plane. Retaining this excess contribution outside the center-accessible interval is essential.

Equation~\eqref{eq:numerics_EL} was solved by logarithmic Picard iteration. The one-dimensional convolutions were evaluated on a uniform grid, and the discrete $W_3$ and $W_2$ kernels were normalized to preserve exactly their zeroth moments, $4\pi R^3/3$ and $4\pi R^2$. The chord stress was evaluated from the distributional formula; in the derivative-expanded implementation the wall delta-functions generated by the jump of $\rho(z)$ were included analytically. The grid spacings were $\Delta z=0.005R$ for the surface-tension and contact-theorem tests, $0.01R$ for the displayed stress profiles, and $0.02R$ for the scan of the disjoining pressure. The convergence criterion for the logarithmic Euler--Lagrange residual was $5\times10^{-10}$ in the first three calculations and $10^{-8}$ in the slit-width scan. Additional implementation details are collected in Appendix~\ref{app:numerics}. At a fixed converged density profile, the thermodynamic surface-free-energy route requires only one-dimensional quadratures and is therefore less expensive than evaluating the chord stress, which adds angular and chord-parameter integrations. For the disjoining pressure the comparison is different: the mechanical route obtains $P_N-p_b$ from a single equilibrium profile, whereas the thermodynamic route requires a numerical derivative with respect to $H$ and therefore profiles at neighboring slit widths. The relative computational cost thus depends on the observable; the chord construction is most useful when spatially resolved stresses or local tractions are required.

\subsection{Hard-wall surface tension}

For a large symmetric slit the two walls are independent to numerical accuracy. We used $H=10\sigma$ and calculated the hard-wall surface free energy by two routes. The thermodynamic route is
\begin{equation}\label{eq:numerics_gamma_th}
 \gamma_{\rm hw}^{\rm th}
 =\frac{1}{2}\left(\frac{\Omega}{A}+p_bH\right)+p_bR,
\end{equation}
where the last term shifts the dividing surface from the contact plane to the physical wall. The mechanical route follows from the finite-slit identity
\begin{equation}\label{eq:numerics_finite_slit_identity}
 I(H)=\int_{-R}^{H+R} \dd z\,[P_N(z)-P_T(z)]
 =\Gamma(H)+H\Pi(H),
\end{equation}
where $\Gamma(H)=\Omega/A+p_bH$ and $\Pi=P_N-p_b$. Appendix~\ref{app:numerics} derives Eq.~\eqref{eq:numerics_finite_slit_identity} by applying the same volume-preserving deformation to the finite-slit grand potential. Hence
\begin{equation}\label{eq:numerics_gamma_mech}
 \gamma_{\rm hw}^{\rm mech}
 =\frac{1}{2}\,[I(H)-H\Pi(H)]+p_bR.
\end{equation}

\begin{figure}[t]
\includegraphics[width=\columnwidth]{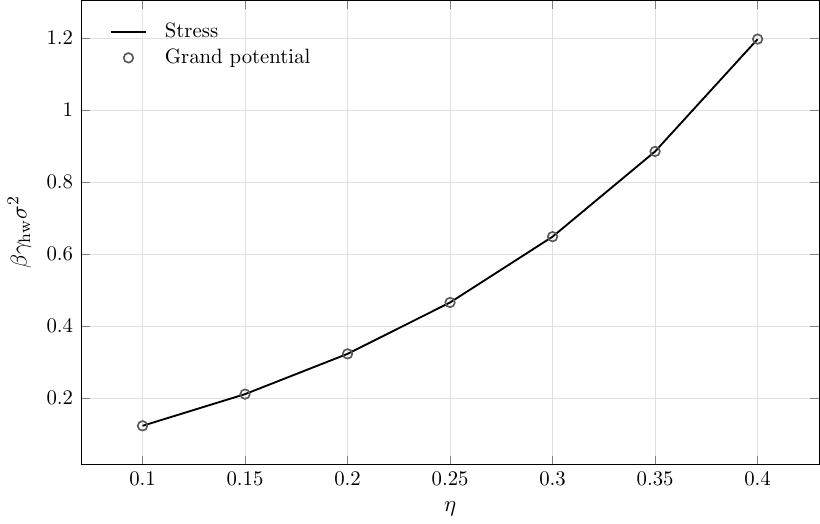}
\caption{Hard-wall surface tension obtained from the thermodynamic excess grand potential and from the mechanical chord-stress route. The plotted quantity is $\beta\gamma_{\rm hw}\sigma^2$. The two routes are independent at the numerical level.}
\label{fig:surface_tension}
\end{figure}

Figure~\ref{fig:surface_tension} shows the result. The two routes are indistinguishable on the scale of the plot. The relative difference is $2.1\times10^{-6}$ at $\eta=0.10$, $1.1\times10^{-5}$ at $\eta=0.20$, $3.9\times10^{-5}$ at $\eta=0.30$, $6.8\times10^{-5}$ at $\eta=0.35$, and $1.1\times10^{-4}$ at $\eta=0.40$. This test is sensitive to both the local $qb$ term and the $\delta'(R-s)$ part of the chord stress.

\subsection{Contact theorem}

For a finite hard slit the contact theorem reads
\begin{equation}\label{eq:numerics_contact}
 \rho(0^+)=\rho(H^-)=\beta P_N,
\end{equation}
where $P_N$ is the normal pressure inside the slit. It is not, in general, equal to the reservoir pressure $p_b$; their difference is the disjoining pressure. We evaluated $P_N$ independently from the chord stress at 31 interior points and used their average. The two contact values were obtained by cubic-spline extrapolation of the converged accessible density profile to $z=0^+$ and $z=H^-$.

\begin{figure}[t]
\includegraphics[width=\columnwidth]{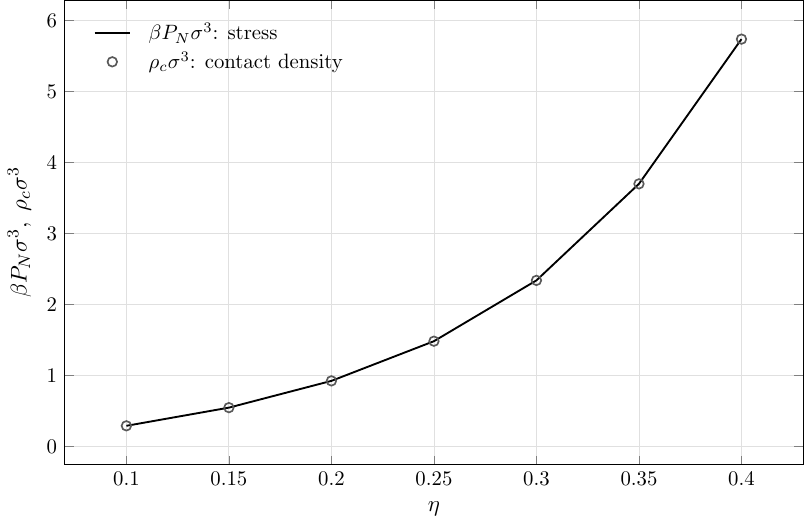}
\caption{Contact theorem in a hard slit of accessible width $H=3\sigma$. The contact density and the normal pressure computed from the chord stress agree over the whole range of packing fractions considered.}
\label{fig:contact}
\end{figure}

Figure~\ref{fig:contact} shows the contact theorem test. The relative errors are $1.2\times10^{-6}$, $9.0\times10^{-6}$, $3.4\times10^{-5}$, and $9.8\times10^{-5}$ for $\eta=0.10$, $0.20$, $0.30$, and $0.40$, respectively. The left and right contact densities agree to machine precision, as expected for a symmetric slit.

\subsection{Normal and tangential stress profiles}

For a slit of accessible width $H=3\sigma$, mechanical equilibrium requires the normal component to be spatially constant in the absence of a body force. The tangential component is not constrained by this condition and reflects the packing oscillations near the walls.

\begin{figure*}[t]
\includegraphics[width=0.92\textwidth]{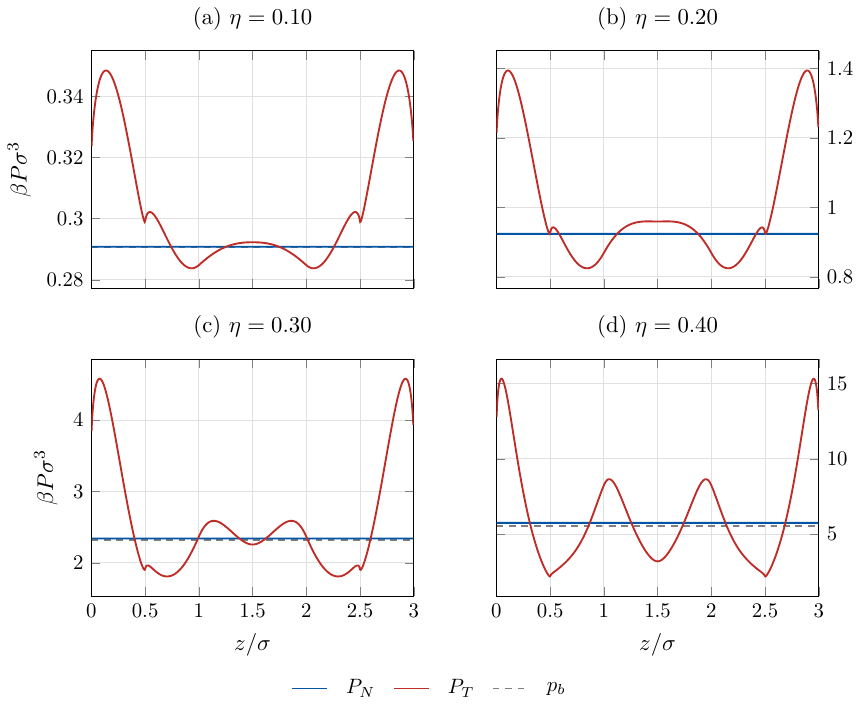}
\caption{Normal and tangential pressure profiles in a hard slit of accessible width $H=3\sigma$. The dashed horizontal line is the reservoir bulk pressure. The normal pressure is constant within the numerical accuracy and differs from the bulk pressure by the disjoining pressure. The tangential pressure is strongly structured near the walls.}
\label{fig:stress_profiles}
\end{figure*}

Figure~\ref{fig:stress_profiles} shows $P_N(z)$ and $P_T(z)$ for four packing fractions. The normal pressure remains constant across the slit, whereas $P_T(z)$ develops pronounced oscillations. At $\eta=0.40$ the tangential pressure varies over more than a factor of six in the slit, while the standard deviation of $P_N$ over the sampled points is $9.1\times10^{-3}$ in units of $\beta P\sigma^3$, about $1.6\times10^{-3}$ of the mean normal pressure. The constancy of $P_N$ is a local test of Eq.~\eqref{eq:force_balance}; the oscillatory tangential profile is the local mechanical expression of packing near the walls.

\subsection{Disjoining pressure}

Finally we compare the disjoining pressure obtained from the stress tensor with a purely thermodynamic route. The mechanical definition is
\begin{equation}\label{eq:numerics_Pi_mech}
 \Pi_{\rm mech}(H)=P_N(H)-p_b.
\end{equation}
The thermodynamic definition is
\begin{equation}\label{eq:numerics_Pi_th}
 \Pi_{\rm th}(H)
 =-\frac{\partial}{\partial H}
 \left[\frac{\Omega(H)}{A}+p_bH\right].
\end{equation}
The derivative was evaluated from a cubic spline of $\Omega/A+p_bH$. The slit widths covered $2\le H/\sigma\le5$ with spacing $\Delta H=0.1R=0.05\sigma$; one additional point was included beyond each end of the displayed interval before differentiating the spline.

\begin{figure*}[t]
\includegraphics[width=0.92\textwidth]{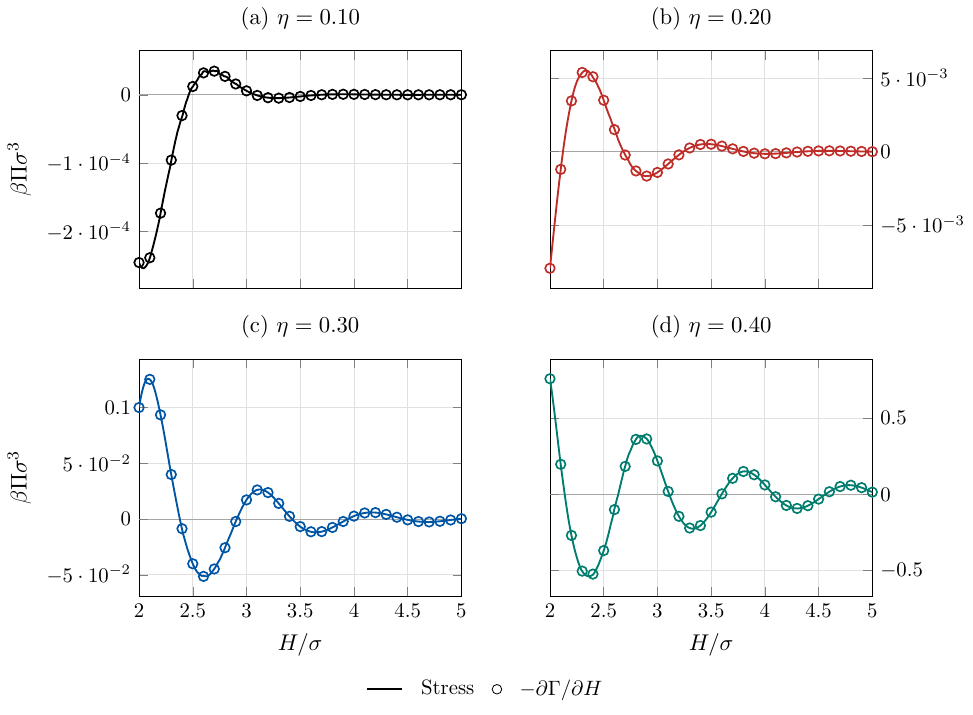}
\caption{Disjoining pressure in a hard slit. Lines show the mechanical route $\Pi_{\rm mech}=P_N-p_b$ obtained from the chord stress. Symbols show the thermodynamic route $\Pi_{\rm th}=-\partial(\Omega/A+p_bH)/\partial H$.}
\label{fig:disjoining}
\end{figure*}

Figure~\ref{fig:disjoining} shows agreement of the two routes throughout the interval $2\le H/\sigma\le5$. The root-mean-square absolute errors in $\beta\Pi\sigma^3$ are $1.3\times10^{-6}$, $1.5\times10^{-5}$, $1.2\times10^{-4}$, and $8.9\times10^{-4}$ for $\eta=0.10$, $0.20$, $0.30$, and $0.40$, respectively. At $H=3\sigma$ the disjoining pressures are weak at low density and become clearly positive at high density; for example, at $\eta=0.40$ we find $\beta\Pi_{\rm mech}\sigma^3=0.2207$ and $\beta\Pi_{\rm th}\sigma^3=0.2196$.

Agreement of the surface-tension, contact, local-balance, and disjoining-pressure tests constrains both the local anisotropic term and the distributional chord contribution.

\section{Discussion and conclusions}
\label{sec:discussion}

An infinitesimal deformation of physical Euclidean space gives a symmetric pressure tensor for the original Rosenfeld fundamental-measure functional. The tensor depends only on the one-body density and the standard FMT weighted fields. Its local part contains the ideal contribution, the excess free-energy density, and the anisotropic term proportional to $n_{2i}n_{2j}$; the nonlocal part distributes the variation of a fundamental measure along the straight segment joining the two points of the convolution.

The finite-range chord term originates from the variation of the Euclidean distance in the FMT kernels; no additional microscopic force-contour prescription is introduced. The straight segment selects one representative of the stress gauge class. Other representatives may differ by identically divergence-free stress modifications, conventionally termed improvement terms in this context; the word ``improvement'' does not imply a more accurate stress tensor. Such modifications leave the bulk pressure, planar normal pressure, surface tension, and disjoining pressure unchanged.

The homogeneous limit gives the Percus--Yevick compressibility pressure of the Rosenfeld functional, and the same tensor obeys the local balance equation in an external field. In planar confinement $P_N$ is spatially constant and satisfies the contact theorem; the integrated anisotropy reproduces the thermodynamic surface excess; and $P_N-p_b$ agrees with the derivative of the confined grand potential. The hard-wall tests require the distributional contribution generated by the density jump and constrain the sign and normalization of the $\delta'(R-s)$ shell term.

The long-wavelength expansion gives an additional check on the finite-range structure of the functional. The bulk quadratic kernel has $\mathcal A(k)=A_0+K_2k^2+K_4k^4+\cdots$, with $K_2<0$ and $K_4>0$ for $0<\eta<1$. The fourth-order truncation is formally stabilized, but its minimum occurs at $kR=O(1)$, outside the controlled regime $kR\ll1$. The packing-scale response encoded by Rosenfeld FMT therefore cannot be represented quantitatively by a low-order Korteweg functional. Appendix~\ref{app:gradient_expansion} gives the explicit coefficients and the dilute second-virial check.

The derivation uses the original monodisperse Rosenfeld functional. Mixtures require species-resolved radii and weighted densities. Tensorial and White-Bear functionals require variation of the additional tensor measures and the modified local free-energy density. In each case, varying a finite-distance kernel produces a distributed chord response, so the same procedure applies to other weighted-density functionals at the one-body level. The practical advantage of the tensorial route appears when a scalar derivative of the grand potential is insufficient. In nonplanar confinement it provides the spatially resolved fluid traction on a boundary and therefore the local mechanical input needed to couple an FMT description of the confined fluid to a deformable solid. The same construction can be extended to mixtures and more elaborate FMT functionals when local mechanical information is required.

\textbf{Data availability statement.} {\sl The data supporting the findings of this study are available in the article.}

\begin{acknowledgments}
This work was supported by the Russian Science Foundation (Grant No. 24-11-00096).
\end{acknowledgments}

\appendix

\section{Contour freedom of the chord representation}
\label{app:general_chord}

The straight segment in Eq.~\eqref{eq:general_chord_stress} follows from Euclidean interpolation between the endpoints of a finite-distance kernel. An alternative path changes the local stress representation without changing its divergence. For any smooth contour $C$ joining $\mathbf r'$ to $\mathbf r$,
\begin{equation}\label{eq:A_general_contour}
 u_i(\mathbf r)-u_i(\mathbf r')
 =
 \int_{C(\mathbf r',\mathbf r)}\dd\ell_j\,\partial_j u_i(\boldsymbol\ell).
\end{equation}
Using Eq.~\eqref{eq:A_general_contour} in place of the straight interpolation produces a corresponding contour representation of the local stress. The associated line distribution satisfies
\begin{equation}\label{eq:A_line_divergence}
 \partial_{z_j}
 \int_C\dd\ell_j\,
 \delta^{(3)}(\mathbf z-\boldsymbol\ell)
 =
 \delta^{(3)}(\mathbf z-\mathbf r')
 -\delta^{(3)}(\mathbf z-\mathbf r).
\end{equation}
The divergence depends only on the endpoints, not on the shape of $C$. If $C_1$ and $C_2$ connect the same endpoints, the difference between the associated line distributions obeys
\begin{equation}\label{eq:A_gauge_divfree}
 \partial_{z_j}
 \left[
 \int_{C_1}\dd\ell_j\,\delta^{(3)}(\mathbf z-\boldsymbol\ell)
 -
 \int_{C_2}\dd\ell_j\,\delta^{(3)}(\mathbf z-\boldsymbol\ell)
 \right]=0.
\end{equation}
The corresponding local pressure tensors therefore differ by a divergence-free contribution. These alternative contours are gauge-related representations of the same local force balance rather than distinct thermodynamic responses. Under the usual regularity and decay conditions this difference may be represented as an improvement term, where ``improvement'' denotes an identically divergence-free modification and does not imply greater physical accuracy,
\begin{equation}\label{eq:A_superpotential}
 \Delta P_{ij}=\partial_k\mathcal A_{ijk},
 \qquad
 \mathcal A_{ijk}=-\mathcal A_{ikj},
\end{equation}
so that $\partial_j\Delta P_{ij}=0$ identically. This is the continuum form of the contour/gauge freedom discussed in microscopic pressure-tensor theory \cite{SchofieldHenderson1982,ShiSmithSantisoGubbins2023}.

The straight chord generated by Euclidean interpolation retains explicit symmetry through the factor $e_i e_j$. It is also the representative selected directly by the variation of the Euclidean distance in the isotropic FMT kernel. Alternative admissible contours preserve the divergence, the bulk pressure, the planar normal pressure, and the integrated planar surface tension, but selecting a different contour as a preferred physical representative would require additional microscopic or constitutive information not contained in the one-body FMT functional.

\section{Reduction of the Rosenfeld measures and vector-weight variation}
\label{app:rosenfeld_vector}

\subsection{Differential hierarchy of the fundamental measures}

Starting from
\begin{equation}\label{eq:B_n3}
 n_3(\mathbf r)
 =
 \int \dd^3r'\,\rho(\mathbf r')\Theta(R-s),
\end{equation}
differentiation with respect to $R$ gives
\begin{equation}\label{eq:B_n2}
 \partial_R n_3(\mathbf r)
 =
 \int \dd^3r'\,\rho(\mathbf r')\delta(R-s)
 =n_2(\mathbf r).
\end{equation}
Similarly,
\begin{equation}\label{eq:B_q}
\begin{aligned}
 \partial_i n_3(\mathbf r)
 &=-\int \dd^3r'\,\rho(\mathbf r')e_i\delta(R-s)\\
 &=-n_{2i}(\mathbf r),
\end{aligned}
\end{equation}
which proves $\mathbf n_2=-\boldsymbol\nabla n_3$. The remaining monodisperse measures differ only by constant geometric factors, giving Eq.~\eqref{eq:dependent_measures}.

Substitution into the original Rosenfeld density \eqref{eq:R_Phi_original} yields Eq.~\eqref{eq:R_Phi_reduced}. Differentiation of that expression gives Eqs.~\eqref{eq:a_explicit}--\eqref{eq:b_explicit}. These formulas were also checked by differentiating the unreduced Rosenfeld functional first and applying the geometrical relations only afterward; the two procedures coincide.

\subsection{Direct variation of the vector weight}

Consider the explicit deformation of
\begin{equation}\label{eq:B_q_direct}
 q_a(\mathbf r)
 =
 \int \dd^3r'\,\rho(\mathbf r')e_a\delta(R-s).
\end{equation}
The variation of the unit vector is
\begin{equation}\label{eq:B_delta_e}
 \delta e_a
 =
 \frac{\delta_{ab}-e_ae_b}{s}
 [u_b(\mathbf r)-u_b(\mathbf r')],
\end{equation}
and
\begin{equation}\label{eq:B_delta_delta}
 \delta\delta(R-s)
 =
 -\delta'(R-s)\delta S.
\end{equation}
Therefore
\begin{equation}\label{eq:B_delta_q_direct}
\begin{aligned}
 \delta q_a(\mathbf r)
 ={}&
 \int \dd^3r'\,\rho(\mathbf r')
 \Bigg\{
 \frac{\delta_{ab}-e_ae_b}{s}\\
 &\qquad\times
 [u_b(\mathbf r)-u_b(\mathbf r')]
 \delta(R-s)\\
 &\qquad
 -e_a\delta'(R-s)\delta S
 \Bigg\}.
\end{aligned}
\end{equation}
The two tensor structures in Eq.~\eqref{eq:B_delta_q_direct} are generated by the derivative identity
\begin{equation}\label{eq:B_kernel_derivative}
 \partial_i[e_j\delta(R-s)]
 =
 \frac{\delta_{ij}-e_i e_j}{s}\delta(R-s)
 -e_i e_j\delta'(R-s).
\end{equation}
Using the chord representation for the displacement difference and integrating by parts with respect to the FMT center coordinate, contraction of Eq.~\eqref{eq:B_delta_q_direct} with $b_a$ becomes
\begin{equation}\label{eq:B_direct_equivalence}
 \int \dd^3r\,b_a\delta q_a
 =
 -\int \dd^3r\,b_iq_jE_{ij}
 +\int \dd^3r\,(\partial_i b_i)\delta n_3.
\end{equation}
Equation~\eqref{eq:B_direct_equivalence} coincides with the result from the transformation law \eqref{eq:q_transform}; generating the vector measure from $n_3$ is therefore equivalent to varying the vector kernel independently.

\section{Planar reduction, distributional terms, and surface tension}
\label{app:planar}

\subsection{Reduction of the three-dimensional kernels}

Let $h=z-z'$ and $r_\perp$ be the transverse separation. For the volume kernel,
\begin{equation}\label{eq:C_W3_derivation}
\begin{aligned}
 W_3(h)
 &=2\pi\int_0^\infty r_\perp\dd r_\perp\,
 \Theta(R-\sqrt{r_\perp^2+h^2})\\
 &=\pi(R^2-h^2)\Theta(R-|h|).
\end{aligned}
\end{equation}
For the surface kernel, changing from $r_\perp$ to $s=\sqrt{r_\perp^2+h^2}$ gives
\begin{equation}\label{eq:C_W2_derivation}
\begin{aligned}
 W_2(h)
 &=2\pi\int_0^\infty r_\perp\dd r_\perp\,
 \delta(R-\sqrt{r_\perp^2+h^2})\\
 &=2\pi\int_{|h|}^{\infty}s\dd s\,\delta(R-s)\\
 &=2\pi R\Theta(R-|h|).
\end{aligned}
\end{equation}
The remaining scalar kernels follow by constant factors. For the $z$ component of the vector surface weight, $e_z=h/s$, so on the shell $s=R$,
\begin{equation}\label{eq:C_V2_derivation}
 V_2(h)=\frac{h}{R}W_2(h)=2\pi h\Theta(R-|h|).
\end{equation}
These kernels satisfy
\begin{equation}\label{eq:C_kernel_norms}
\begin{aligned}
 \int_{-\infty}^{\infty}\dd h\,W_0&=1,
 &
 \int_{-\infty}^{\infty}\dd h\,W_1&=R,\\
 \int_{-\infty}^{\infty}\dd h\,W_2&=4\pi R^2,
 &
 \int_{-\infty}^{\infty}\dd h\,W_3&=\frac{4\pi R^3}{3}.
\end{aligned}
\end{equation}
and
\begin{equation}\label{eq:C_vector_norms}
 \int_{-\infty}^{\infty}\dd h\,V_1
 =\int_{-\infty}^{\infty}\dd h\,V_2=0.
\end{equation}

\subsection{Planar chord components}

Starting from Eq.~\eqref{eq:C_one_center}, write $\dd^3s=s^2\dd s\,\dd\Omega$ and $u=\cos\theta$. The normal projection is $e_z^2=u^2$. Azimuthal integration gives
\begin{equation}\label{eq:C_Czz_dist}
\begin{aligned}
 C_{zz}(z)
 ={}&
 2\pi\int_0^1\dd\lambda\int_{-1}^{1}\dd u
 \int_0^\infty\dd s\,
 s^3u^2\rho(z-\lambda su)\\
 &\times
 \Big\{
 A[z+(1-\lambda)su]\delta(R-s)\\
 &\qquad+
 \chi[z+(1-\lambda)su]\delta'(R-s)
 \Big\}.
\end{aligned}
\end{equation}
For one tangential component, the azimuthal average of $e_x^2$ is $(1-u^2)/2$, and therefore
\begin{equation}\label{eq:C_Cxx_dist}
\begin{aligned}
 C_{xx}(z)
 ={}&
 \pi\int_0^1\dd\lambda\int_{-1}^{1}\dd u
 \int_0^\infty\dd s\,
 s^3(1-u^2)\rho(z-\lambda su)\\
 &\times
 \Big\{
 A[z+(1-\lambda)su]\delta(R-s)\\
 &\qquad+
 \chi[z+(1-\lambda)su]\delta'(R-s)
 \Big\}.
\end{aligned}
\end{equation}
Equations~\eqref{eq:C_Czz_dist} and \eqref{eq:C_Cxx_dist} are valid distributionally and are the appropriate formulas when a hard wall generates a discontinuity in $\rho(z)$.

\subsection{Action of the \texorpdfstring{$\delta'$}{delta-prime} shell}

For a smooth test function $f(s)$,
\begin{equation}\label{eq:C_delta_prime_proof1}
 \int_0^\infty\dd s\,f(s)\delta'(R-s)
 =f'(R).
\end{equation}
This follows because $\delta'(R-s)=-\delta'(s-R)$ and the standard distribution identity is $\int f(s)\delta'(s-R)\dd s=-f'(R)$. In Eq.~\eqref{eq:C_delta_prime_proof1}, the prime on $f$ means differentiation with respect to the integration variable $s$. Any dependence of the FMT fields on the physical hard-sphere radius is held fixed during this operation.

For the present problem,
\begin{equation}\label{eq:C_test_f}
\begin{aligned}
 f(s)={}&s^3\rho(z-\lambda su)
 \chi[z+(1-\lambda)su],
\end{aligned}
\end{equation}
so
\begin{equation}\label{eq:C_fprime}
\begin{aligned}
 f'(R)={}&
 3R^2\rho_-\chi_+
 -\lambda R^3u\,\rho_-'\chi_+\\
 &+(1-\lambda)R^3u\,\rho_-\chi_+',
\end{aligned}
\end{equation}
where spatial primes denote derivatives with respect to the $z$ coordinate at the indicated arguments. Substitution into Eq.~\eqref{eq:G_def} gives Eqs.~\eqref{eq:Czz} and \eqref{eq:Cxx} for a smooth profile.

For a hard wall at the center-accessible plane $z=0$, write
\begin{equation}\label{eq:C_rho_step}
 \rho(z)=\Theta(z)\rho_+(z).
\end{equation}
Then
\begin{equation}\label{eq:C_wall_derivative}
\begin{aligned}
 \frac{\partial}{\partial s}\rho(z-\lambda su)
 ={}&-\lambda u\Big\{
 \Theta(z-\lambda su)\rho_+'(z-\lambda su)\\
 &\hspace{18mm}
 +\delta(z-\lambda su)\rho_+(0^+)
 \Big\}.
\end{aligned}
\end{equation}
The second term is a genuine surface contribution. Omitting it by differentiating only the smooth interior profile changes the mechanical surface excess and destroys the contact/surface-tension consistency. For a slit, write
\begin{equation}\label{eq:C_slit_rho}
 \rho(z)=\Theta(z)\Theta(H-z)\rho_+(z).
\end{equation}
With $x=z-\lambda su$,
\begin{equation}\label{eq:C_slit_wall_derivative}
\begin{aligned}
 \frac{\partial}{\partial s}\rho(x)
 ={}&-\lambda u\Big\{
 \Theta(x)\Theta(H-x)\rho_+'(x)\\
 &\qquad+\delta(x)\rho_+(0^+)
 -\delta(H-x)\rho_+(H^-)
 \Big\}.
\end{aligned}
\end{equation}
The opposite signs of the two delta-functions reflect the opposite orientations of the two contact planes. In numerical work one may either retain Eqs.~\eqref{eq:C_Czz_dist}--\eqref{eq:C_Cxx_dist} directly or evaluate Eqs.~\eqref{eq:C_wall_derivative} and \eqref{eq:C_slit_wall_derivative} analytically together with the regular part.

Particle centers are excluded from $z<0$, but weighted densities centered at $-R<z<0$ remain nonzero because their integration spheres overlap the accessible fluid. Hence
\begin{equation}\label{eq:C_outside_Phi}
 \Phi(z)\neq0\quad\text{in general for}\quad -R<z<0.
\end{equation}
All thermodynamic and deformation integrals must include the full support of these weighted fields.

\subsection{Integrated stress anisotropy}

From Eqs.~\eqref{eq:C_Czz_dist} and \eqref{eq:C_Cxx_dist},
\begin{equation}\label{eq:C_gamma_start}
\begin{aligned}
 \int_{-\infty}^{\infty}\dd z\,[C_{zz}-C_{xx}]
 ={}&
 \pi\int_0^1\dd\lambda\int_{-1}^{1}\dd u\,
 (3u^2-1)\\
 &\times\int_{-\infty}^{\infty}\dd z\int_0^\infty\dd s\,
 s^3\rho(z-\lambda su)\\
 &\times
 \Big\{
 A[z+(1-\lambda)su]\delta(R-s)\\
 &\qquad+
 \chi[z+(1-\lambda)su]\delta'(R-s)
 \Big\}.
\end{aligned}
\end{equation}
Set $z_p=z-\lambda su$. Then $\dd z=\dd z_p$ and
\begin{equation}\label{eq:C_gamma_endpoint}
 z+(1-\lambda)su=z_p+su.
\end{equation}
The integrand becomes independent of $\lambda$, so the chord-parameter integral equals unity. For smooth profiles, performing the $s$ integration then gives
\begin{equation}\label{eq:C_gamma_chord_final}
\begin{aligned}
 \int_{-\infty}^{\infty}\dd z\,[C_{zz}-C_{xx}]
 ={}&
 \pi\int_{-1}^{1}\dd u\,(3u^2-1)\\
 &\times
 \int_{-\infty}^{\infty}\dd z\,\rho(z)
 \Big[
 R^3A(z+Ru)\\
 &\hspace{17mm}
 +3R^2\chi(z+Ru)\\
 &\hspace{17mm}
 +R^3u\chi'(z+Ru)
 \Big].
\end{aligned}
\end{equation}
Adding the local anisotropic contribution $\int_{-\infty}^{\infty}q(z)b(z)\,\dd z$ yields Eq.~\eqref{eq:gamma_compact}. For a hard wall, Eq.~\eqref{eq:C_gamma_chord_final} is obtained most safely by performing the $z$ integration in the distributional expression \eqref{eq:C_gamma_start} before the radial $s$ integration. After the substitution $z_p=z-\lambda su$, $\rho(z_p)$ is independent of $s$; the $\delta'(R-s)$ shell therefore differentiates only $s^3\chi(z_p+su)$. The wall jump is then already accounted for by the integration domain, and no derivative of the discontinuous density is needed. If the local derivative-expanded form is used instead, the surface term in Eq.~\eqref{eq:C_wall_derivative} must be kept explicitly.

\section{Long-wavelength expansion of the Rosenfeld functional}
\label{app:gradient_expansion}

The local gradient limit is determined from the bulk quadratic response of the full nonlocal functional. The expansion below tests whether a finite number of gradient terms remains controlled on the packing scale.

\subsection{Bulk quadratic kernel}

Let
\begin{equation}\label{eq:E_density_fluctuation}
 \rho(\mathbf r)=\rho_b+\delta\rho(\mathbf r),
 \qquad
 \rho_b=\frac{3\eta}{4\pi R^3}.
\end{equation}
To quadratic order in the density fluctuation,
\begin{equation}\label{eq:E_quadratic_free_energy}
 \delta^2\beta F
 =
 \frac{1}{2}
 \int\frac{\dd^3k}{(2\pi)^3}\,
 \mathcal A(k)|\delta\rho_{\mathbf k}|^2.
\end{equation}
The Fourier transforms of the scalar volume and surface weights are
\begin{equation}\label{eq:E_w3_fourier}
 \widetilde w_3(k)
 =
 4\pi R^3
 \frac{\sin x-x\cos x}{x^3},
 \qquad x=kR,
\end{equation}
\begin{equation}\label{eq:E_w2_fourier}
 \widetilde w_2(k)
 =
 4\pi R^2\frac{\sin x}{x}.
\end{equation}
Since $\mathbf q=-\boldsymbol\nabla n_3$, the vector fluctuation satisfies
\begin{equation}\label{eq:E_q_fourier}
 \delta\mathbf q_{\mathbf k}
 =
 -i\mathbf k\,\widetilde w_3(k)\delta\rho_{\mathbf k}.
\end{equation}
Define the bulk derivatives
\begin{equation}\label{eq:E_Phi_derivatives_def}
 \Phi_{33}=\frac{\partial^2\Phi}{\partial n_3^2},
 \qquad
 \Phi_{32}=\frac{\partial^2\Phi}{\partial n_3\partial n_2},
 \qquad
 \Phi_{22}=\frac{\partial^2\Phi}{\partial n_2^2},
 \qquad
 \Phi_q=\frac{\partial\Phi}{\partial q^2},
\end{equation}
all evaluated at $n_3=\eta$, $n_2=3\eta/R$, and $q=0$. The quadratic kernel generated by the Rosenfeld functional is then
\begin{equation}\label{eq:E_exact_kernel}
\begin{aligned}
 \mathcal A(k)={}&
 \frac{1}{\rho_b}
 +\Phi_{33}\widetilde w_3^2(k)
 +2\Phi_{32}\widetilde w_3(k)\widetilde w_2(k)\\
 &+\Phi_{22}\widetilde w_2^2(k)
 +2\Phi_q k^2\widetilde w_3^2(k).
\end{aligned}
\end{equation}
For the original Rosenfeld density, direct differentiation gives
\begin{equation}\label{eq:E_bulk_Hessian}
\begin{aligned}
 \Phi_{33}&=
 \frac{3\eta(1+2\eta)^2}
 {4\pi R^3(1-\eta)^4},\\
 \Phi_{32}&=
 \frac{(1+2\eta)^2}
 {4\pi R^2(1-\eta)^3},\\
 \Phi_{22}&=
 \frac{\eta+2}
 {4\pi R(1-\eta)^2},\\
 \Phi_q&=
 -\frac{\eta+2}
 {8\pi R(1-\eta)^2}.
\end{aligned}
\end{equation}
Equation~\eqref{eq:E_exact_kernel} is the inverse bulk density response,
\begin{equation}\label{eq:E_structure_factor_relation}
 \mathcal A(k)=\frac{1}{\rho_b S_{\rm FMT}(k)},
\end{equation}
where $S_{\rm FMT}(k)$ is the bulk structure factor generated by the Rosenfeld functional; for the original Rosenfeld functional this bulk response coincides with the Percus--Yevick hard-sphere result. Equation~\eqref{eq:E_exact_kernel} therefore contains the full finite-range information of the FMT weights.

\subsection{Gradient expansion}

For $x=kR\ll1$,
\begin{equation}\label{eq:E_weights_expansion}
\begin{aligned}
 \widetilde w_3(k)
 &=\frac{4\pi R^3}{3}
 \left(1-\frac{x^2}{10}+\frac{x^4}{280}
 -\frac{x^6}{15120}+O(x^8)\right),\\
 \widetilde w_2(k)
 &=4\pi R^2
 \left(1-\frac{x^2}{6}+\frac{x^4}{120}
 -\frac{x^6}{5040}+O(x^8)\right).
\end{aligned}
\end{equation}
Substitution into Eq.~\eqref{eq:E_exact_kernel} gives
\begin{equation}\label{eq:E_kernel_expansion}
 \mathcal A(k)
 =A_0+K_2k^2+K_4k^4+K_6k^6+O(k^8),
\end{equation}
where
\begin{equation}\label{eq:E_A0}
 A_0
 =
 \frac{4\pi R^3(1+2\eta)^2}
 {3\eta(1-\eta)^4},
\end{equation}
\begin{equation}\label{eq:E_K2}
 K_2
 =
 -\frac{4\pi R^5}{15}
 \frac{4\eta^2-11\eta+16}{(1-\eta)^4},
\end{equation}
\begin{equation}\label{eq:E_K4}
 K_4
 =
 \frac{4\pi R^7}{525}
 \frac{7\eta^3+12\eta^2-72\eta+80}
 {(1-\eta)^4},
\end{equation}
\begin{equation}\label{eq:E_K6}
 K_6
 =
 -\frac{16\pi R^9}{14175}
 \frac{6\eta^3+4\eta^2-41\eta+40}
 {(1-\eta)^4}.
\end{equation}
The quadratic polynomial in Eq.~\eqref{eq:E_K2} has negative discriminant, so $K_2<0$ for $0<\eta<1$. The cubic numerator in Eq.~\eqref{eq:E_K4} decreases on this interval and equals $27$ at $\eta=1$, hence $K_4>0$. The numerator in Eq.~\eqref{eq:E_K6} is also positive on $0<\eta<1$, and therefore $K_6<0$.

At order $k^2$, $K_2$ is the square-gradient coefficient of the local Korteweg truncation. Its negative sign makes that second-order theory unstable at sufficiently large wave number. Retaining $K_4$ gives the formally stabilized kernel
\begin{equation}\label{eq:E_quartic_kernel}
 \mathcal A_4(k)=A_0+K_2k^2+K_4k^4,
\end{equation}
whose nonzero stationary point is
\begin{equation}\label{eq:E_kstar}
 (k_*R)^2
 =
 \frac{35(4\eta^2-11\eta+16)}
 {2(7\eta^3+12\eta^2-72\eta+80)}.
\end{equation}
For $0.10\leq\eta\leq0.49$, Eq.~\eqref{eq:E_kstar} gives $1.89\lesssim k_*R\lesssim2.04$. The quartic minimum remains positive, so this truncation does not predict a finite-wave-number instability. Its location, however, is already outside the controlled long-wavelength regime $kR\ll1$.

The unexpanded kernel \eqref{eq:E_exact_kernel}, equivalently the inverse of the FMT structure factor, has its first minima at $kR=2.954$, $3.049$, $3.169$, $3.320$, and $3.483$ for $\eta=0.10$, $0.20$, $0.30$, $0.40$, and $0.49$, respectively. These wave numbers correspond to packing-scale density structure. The alternating signs $K_2<0$, $K_4>0$, and $K_6<0$ show that a finite-order local gradient expansion cannot reproduce this regime quantitatively; the finite-range FMT kernel must be retained.

\subsection{Dilute second-virial limit}

The low-density limit provides an independent check. From Eqs.~\eqref{eq:E_K2}--\eqref{eq:E_K6},
\begin{equation}\label{eq:E_dilute_coefficients}
\begin{aligned}
 K_2&\xrightarrow[\eta\to0]{}
 -\frac{64\pi R^5}{15},\\
 K_4&\xrightarrow[\eta\to0]{}
 \frac{64\pi R^7}{105},\\
 K_6&\xrightarrow[\eta\to0]{}
 -\frac{128\pi R^9}{2835}.
\end{aligned}
\end{equation}
The exact second-virial excess functional of hard spheres is
\begin{equation}\label{eq:E_second_virial}
 \beta F_{\rm ex}^{(2)}
 =
 \frac{1}{2}
 \int\dd^3r\,\dd^3s\,
 \rho(\mathbf r)\rho(\mathbf r-\mathbf s)
 \Theta(2R-s).
\end{equation}
The Fourier transform of its excluded-volume kernel is
\begin{equation}\label{eq:E_second_virial_kernel}
\begin{aligned}
 \widetilde\Theta_{2R}(k)
 =32\pi R^3\Bigg[
 &\frac{1}{3}
 -\frac{(2kR)^2}{30}
 +\frac{(2kR)^4}{840}\\
 &-\frac{(2kR)^6}{45360}
 +O(k^8R^8)
 \Bigg].
\end{aligned}
\end{equation}
Its $k^2$, $k^4$, and $k^6$ coefficients coincide with Eq.~\eqref{eq:E_dilute_coefficients}. The long-wavelength expansion of Rosenfeld FMT therefore reproduces the exact second-virial result through sixth order, while the finite-density packing regime remains intrinsically nonlocal.

\section{Numerical implementation and finite-slit identities}
\label{app:numerics}

The calculations in Sec.~\ref{sec:numerical} use the formulas and conventions specified below.

\subsection{Euler--Lagrange iteration}

For a prescribed reservoir packing fraction $\eta$, the bulk density is
\begin{equation}\label{eq:D_rhob}
 \rho_b=\frac{3\eta}{4\pi R^3}.
\end{equation}
The bulk weighted fields are $n_{3b}=\eta$, $n_{2b}=3\eta/R$, and $q_b=0$. Equations~\eqref{eq:numerics_mu_bulk} and \eqref{eq:PY_EOS} then determine $\mu$ and $p_b$.

For a current profile $\rho^{(m)}(z)$, Eq.~\eqref{eq:numerics_cex} gives $c_{\rm ex}^{(m)}(z)$ and the undamped Euler--Lagrange update is
\begin{equation}\label{eq:D_picard_target}
 \rho_*(z)=\Lambda^{-3}
 \exp\!\left[\beta\mu-c_{\rm ex}^{(m)}(z)\right].
\end{equation}
We used logarithmic mixing,
\begin{equation}\label{eq:D_log_mixing}
 \ln\rho^{(m+1)}
 =(1-\alpha)\ln\rho^{(m)}+\alpha\ln\rho_*,
\end{equation}
with $\alpha$ between $0.08$ and $0.12$ and additional damping if the residual increased. Positivity of the density is then automatic. Convergence was monitored with
\begin{equation}\label{eq:D_EL_residual}
 \mathcal R_{\rm EL}
 =\max_{0<z<H}\left|
 \ln\rho^{(m)}(z)-\ln\rho_*(z)
 \right|.
\end{equation}

\subsection{Discrete planar convolutions}

On a uniform grid $z_j$, Eqs.~\eqref{eq:1D_scalar_conv} and \eqref{eq:1D_vector_conv} were evaluated as discrete convolutions. The scalar kernels $W_3$ and $W_2$ were rescaled by a grid-level constant so that
\begin{equation}\label{eq:D_discrete_norm}
 \Delta z\sum_j W_3(z_j)=\frac{4\pi R^3}{3},
 \qquad
 \Delta z\sum_j W_2(z_j)=4\pi R^2
\end{equation}
exactly. The vector kernel $V_2$ was kept antisymmetric, so its discrete zeroth moment vanishes. At the two grid points corresponding to $|h|=R$, the discontinuous $W_2$ and $V_2$ kernels were assigned the half-weight appropriate to the trapezoidal rule before the scalar normalization was imposed. Without this normalization, a small error in the bulk zeroth moments generates an artificial term proportional to $H$ in $\Omega/A+p_bH$.

The local stress was evaluated from the distributional chord formula. For the derivative-expanded representation, the density was written as a smooth interior profile multiplied by the two hard-wall step functions. The resulting delta-functions at $z=0$ and $z=H$ were integrated analytically. Smooth derivatives of $b(z)$ and $\chi(z)$ were evaluated by cubic splines, while the wall jump itself was never differentiated numerically. The angular and chord-parameter integrals were evaluated by Gauss--Legendre quadrature. For the contact-theorem and displayed stress-profile calculations we used 48 angular nodes and 18 nodes on each smooth $\lambda$ subinterval; the $\lambda$ interval was split whenever a chord first crossed a wall. The integrated surface-tension expression used 160 angular nodes. In the slit-width scan, where only the central normal stress was required, 32 angular nodes and 12 nodes per $\lambda$ subinterval were sufficient at $\Delta z=0.02R$.

\subsection{Contact theorem in a slit}

For a hard wall, an infinitesimal displacement of the contact plane changes the grand potential by the mechanical work $P_N A\,\delta\ell$. The same displacement changes the configurational integral through the loss or gain of an infinitesimal accessible layer; the corresponding change is $k_BT\rho_c A\,\delta\ell$, where $\rho_c$ is the one-sided contact density. Equating the two variations gives
\begin{equation}\label{eq:D_contact_derivation}
 \rho_c=\beta P_N.
\end{equation}
In a symmetric slit, $P_N$ is constant and the two contact densities are equal. Numerically, $P_N$ was averaged over 31 interior points, while the one-sided contact values were obtained from cubic-spline extrapolation of the converged profile.

\subsection{Finite-slit stress identity}

Let
\begin{equation}\label{eq:D_omega_Gamma}
 \omega(H)=\frac{\Omega(H)}{A},
 \qquad
 \Gamma(H)=\omega(H)+p_bH.
\end{equation}
Here $H$ refers to the distance between the center-accessible planes and $R$ is held fixed; the shift to the physical-wall dividing surface is applied only after the contact-plane excess has been obtained. Under the volume-preserving affine deformation \eqref{eq:area_deformation}, $\delta A/A=2\epsilon$ and $\delta H=-2\epsilon H$. Hence
\begin{equation}\label{eq:D_finite_slit_thermo}
\begin{aligned}
 \delta\Omega
 &=2\epsilon A\left[\omega(H)-H\omega'(H)\right]\\
 &=2\epsilon A\left[\Gamma(H)+H\Pi(H)\right],
\end{aligned}
\end{equation}
where
\begin{equation}\label{eq:D_Pi_identity}
 \Pi(H)=-\Gamma'(H)=-\omega'(H)-p_b=P_N-p_b.
\end{equation}
The virtual-work expression for the same deformation is
\begin{equation}\label{eq:D_finite_slit_mech}
 \delta\Omega
 =2\epsilon A
 \int_{-R}^{H+R}\dd z\,[P_N(z)-P_T(z)].
\end{equation}
Equating Eqs.~\eqref{eq:D_finite_slit_thermo} and \eqref{eq:D_finite_slit_mech} proves Eq.~\eqref{eq:numerics_finite_slit_identity}. In the limit of two independent walls, subtracting $H\Pi$ and dividing by two gives the contact-plane surface excess; the final $p_bR$ shift moves the dividing surface to the physical wall.

% Bibliography embedded for arXiv single-source submission
%aipnum4-2.bst 2019-01-14 (MD) hand-edited version of apsrev4-1.bst

\end{document}